\documentclass[10pt,letterpaper]{article}
\usepackage[top=0.85in,left=2.75in,footskip=0.75in]{geometry}

\usepackage{amsmath,amssymb}

\usepackage{changepage}

\usepackage[utf8x]{inputenc}

\usepackage{textcomp,marvosym}

\usepackage{cite}

\usepackage{nameref,hyperref}

\usepackage[right]{lineno}

\usepackage{microtype}
\DisableLigatures[f]{encoding = *, family = * }

\usepackage[table]{xcolor}

\usepackage{array}

\newcolumntype{+}{!{\vrule width 2pt}}

\newlength\savedwidth

\raggedright
\usepackage[aboveskip=1pt,labelfont=bf,labelsep=period,justification=raggedright,singlelinecheck=off]{caption}

\usepackage{xspace} 
\newcommand{\etal}{\textit{et al.}\@\xspace }
\newcommand{\eg}{e.g.\@\xspace}

\makeatletter
\renewcommand{\@biblabel}[1]{\quad#1.}
\makeatother

\usepackage{lastpage,fancyhdr,graphicx}
\usepackage{epstopdf}
\fancyheadoffset[L]{2.25in}
\fancyfootoffset[L]{2.25in}
\begin{document}
\vspace*{0.2in}

\begin{flushleft}
{\Large
\textbf\newline{Modelling COVID-19 transmission in supermarkets using an agent-based model}
}
\newline
\\
Fabian Ying \textsuperscript{1},
Neave O'Clery\textsuperscript{2*},
\\
\bigskip
\textbf{1} G-Research, London, UK
\\
\textbf{2} Centre for Advanced Spatial Analysis, UCL, London, UK
\\
\bigskip

%
%





* n.oclery@ucl.ac.uk

\end{flushleft}
\section*{Abstract}
Since the outbreak of COVID-19 in early March 2020, supermarkets around the world have implemented different policies to reduce the virus transmission in stores to protect both customers and staff, such as restricting the maximum number of customers in a store, changes to the store layout, or enforcing a mandatory face covering policy. 
To quantitatively assess these mitigation methods, we formulate an agent-based model of customer movement in a supermarket (which we represent by a network) with a simple virus transmission model based on the amount of time a customer spends in close proximity to infectious customers. 
We apply our model to synthetic store and shopping data to show how one can use our model to estimate the number of infections due to human-to-human contact in stores and how to model different store interventions.
The source code is openly available under \href{https://github.com/fabianying/covid19-supermarket-abm}{https://github.com/fabianying/covid19-supermarket-abm}.
We encourage retailers to use the model to find the most effective store policies that reduce virus transmission in stores and thereby protect both customers and staff.


\section*{Introduction}
    As the main provider of food and essential goods, supermarkets remained open in many countries throughout the COVID-19 pandemic in 2020, while the majority of other businesses (such as general retail stores) shut down during periods of government-mandated lockdowns \cite{politico_lockdown_comparison,hale2020oxford}. 
    Supermarkets represent one of the main hubs where a large number of people mix indoors throughout the pandemic and are thus a potential risk area where the virus SARS-CoV-2, which causes COVID-19, may be transmitted.
    It is therefore vital to find safe ways for customers to shop and minimize virus transmission. 
    Models for customer dynamics and virus transmission are useful towards that goal, as they can be used to estimate the infection risk and assess how different interventions affect the risk.
    In this article, we propose an agent-based model for customer dynamics which we use to estimate the total amount of \emph{exposure time}, which we define as the amount of time that customers are in close proximity to infected customers.
    Using a simple virus transmission model, we estimate the number of infections from the exposure time.
    We apply this model to synthetic data and how to model the following interventions:
    \begin{itemize}
         \item Controlling the rate of customer arrival,
         \item Restricting the maximum number of customers in the store,
         \item Implementing face mask policy, and
         \item One-way aisle store layout.
     \end{itemize}
     These and other interventions have been used or recommended in supermarkets in the UK and the US \cite{grocer2020lockdownmeasures,cdc2020covid19supermarketadvice,ukgov2020covidsupermarketguidance}, among other countries.

\section*{Materials and methods}
    \label{sec:materials_and_methods}

    \subsection*{Store graph}
\begin{figure}[!htb]
        \centering
        \includegraphics[width=\textwidth,trim={6cm 1cm 6cm 4cm},clip]{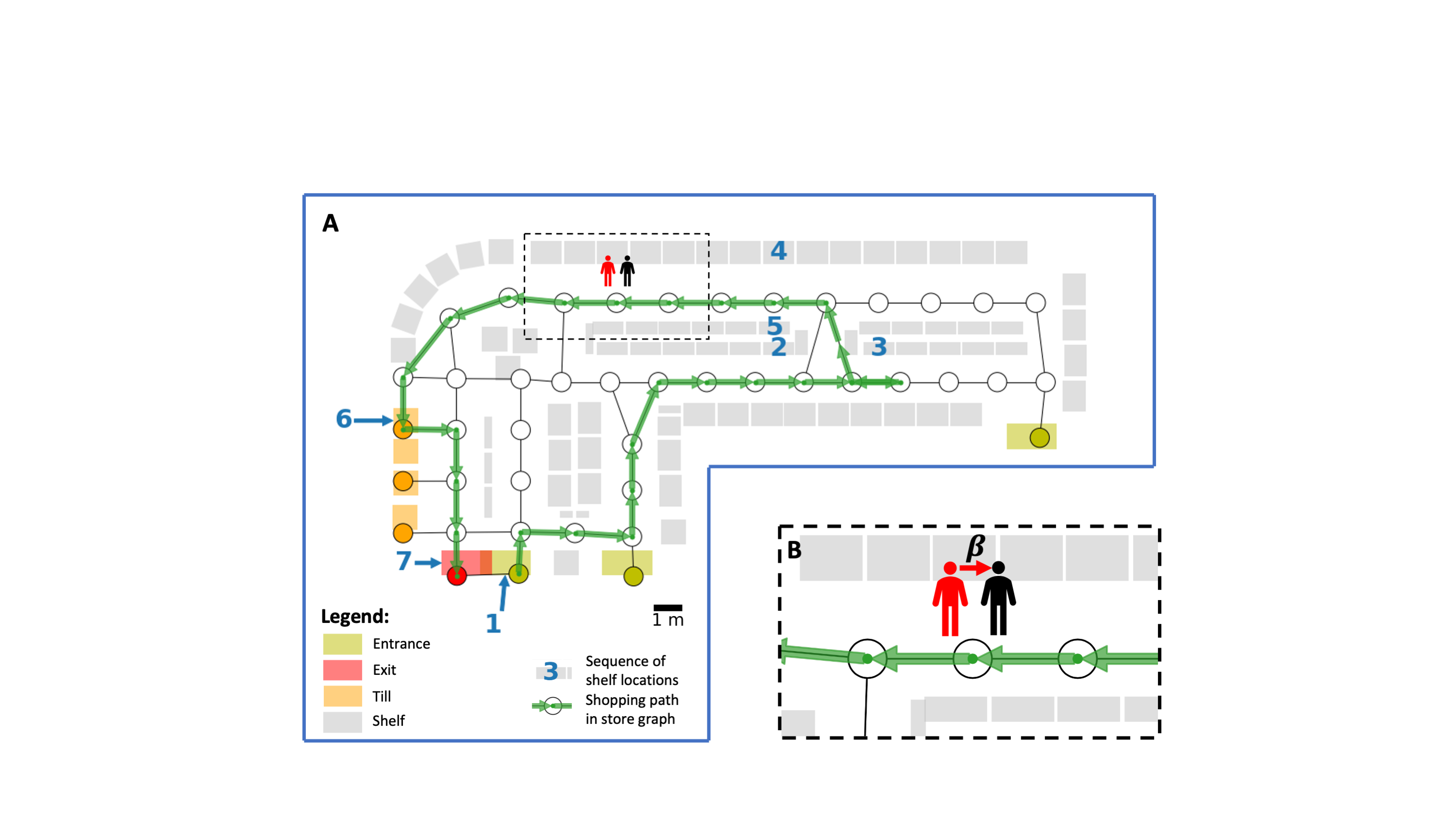}
        \caption{
        \textbf{(A)} A network representation of a small supermarket/convenience store with an example shopping path (in green). 
        We generate each shopping path from a sequence of shelf locations (in blue), which correspond to the shelves from a customer picks up their items during a visit and the entrance and the tills.
        \textbf{(B)} Virus transmission model. A susceptible customer (in black) becomes infected at rate $\beta$ whenever they are in the same zone as an infectious customer (in red).}
        \label{fig:layout}
    \end{figure}

    We represent a store as a network (called a \emph{store graph}), in which nodes represent zones and edges connect contiguous zones.
    We create a store graph from a synthetic store layout following a similar procedure as in \cite{Ying2019,ying2019PhDThesis}. 
    Zones are approximately 2m by 2m and we specify a number of entrance, till, and exit nodes (see Fig.~\ref{fig:layout}A).

    \subsection*{Agent-based model}
        Our agent-based model has two components: a customer mobility model and a virus transmission model.

        \subsubsection*{Customer mobility model}
            \label{sub:customermovement}
            In our agent-based model, customers arrive the store according to a Poisson process with constant rate $\lambda$. 
            Each customer starts at a random entrance node (chosen uniformly at random from all entrance nodes) and is assigned a shopping path, chosen uniformly at random from all shopping paths from an empirical or synthetic data set. 
            Each shopping path is a path in the store graph, representing the route that a customer takes in the store.
            Two consecutive nodes in the shopping path may be identical.
            This case occurs when a customer picks up one or more items in the zone.
            A customer traverses the store graph according to its assigned shopping path. 
            At each node, a customer waits a random time $T$, which is exponentially distributed with mean $\tau$ (independent of other waiting times), before traversing to the next node in the shopping path (or staying at the same node, if it is the next node).
            Once a customer arrives at the final node (which is the exit node) in its shopping path, the customer stays $T$ seconds on the node (with $T$ again exponentially distributed with mean $\tau$) and is then removed from the system.
            At the beginning of each simulation, the store is empty and customers arrive in the store over a period of $H$ hours (corresponding to length of the opening hours of the store). After $H$ hours, no new customers arrive and the simulation stops once the last customer leaves the store.
    
        \subsubsection*{Virus transmission model}
            \label{sub:infectionmechanism}
            Customer are either susceptible or infectious when they enter the store.
            Each customer that arrives to the store is infectious with independent probability $p_{I}$ (corresponding to the proportion of infectious customers) and is otherwise susceptible.
            In our infection mechanism, we assume susceptible customers become infected proportional to the time they spent with infectious customers. 
            More formally, we define the \emph{total exposure time} $E_s$ for each susceptible customer $s$ as follows.
            We define the \emph{individual exposure time} $E_s^{(i)}$ to an infectious customer $i$ as the total time that customer $s$ was in the same zone as an infectious customer $i$ during the shopping trip of $s$.
            Then $E_s = \sum_{i} E_s^{(i)}$ is the sum of the individual exposure times.
            Each susceptible customer becomes infected after the shopping trip with probability $\min(\beta E_s, 1)$ for some transmission parameter $\beta > 0$.
            We illustrate the virus transmission model in see Fig.~\ref{fig:layout}B.

    \subsection*{Data}
        \label{sec:data}
        We use a synthetically created store layout and shopping path data.
        The (synthetic) store is a small store with around 80 shelves, four tills, three entrances, and one exit (see Fig~\ref{fig:layout}).
        We synthetically generate shopping paths as follows. 
        For each path, we create a sequence of locations $(s_1, \dots, s_k)$, where $s_1$ is a random entrance, $s_2, \dots, s_{k-1}$ are random shelves, and $s_k$ is a random till.
        The locations $s_2,\dots,s_{k-1}$ represent the shelves where a customer buys their items in their shopping trip.
        We map each location $s_i$ to the corresponding node $v_i$ in the store graph that contains the shelf, entrance, or exit location.
        We then generate the shopping path from the node sequence $(v_1, \dots, v_k)$ by choosing a shortest path in the store graph that visits each of these nodes in sequence.
        We show an example sequence of locations and its corresponding shopping path in Fig~\ref{fig:layout}A.

    \subsection*{Parameter values}
        \newcommand{\betaVal}{\ensuremath{1.41 \times 10^{-9}}}
        \newcommand{\pIVal}{\ensuremath{0.11\%}}
        \newcommand{\meanWaitTime}{\ensuremath{0.2}}

        In our simulations, we set the default arrival rate to be 2.55 customers per minute. 
        This is based on the mean number of baskets per store over a 91-day period across 17 UK supermarkets as reported in \cite{Ying2019} and the typical UK store opening period of 14 hours \cite{thesunsupermarketopeningtimes}.
        (In \cite{Ying2019}, Ying \etal used a data set of 13672 mean baskets per store over 91 days, representing a sample of 7\% of the total number of baskets. Therefore, the customer arrival rate is $13672 / (0.07 \times 91 \times 14 \times 60) = 2.55$ customers per minute over this period.)
        We assume that each basket corresponds to a single customer (rather than groups of customers).
        At the time of writing, many UK supermarkets advise or restrict customers to shop alone \cite{lovemoneyhouseholdshopping}.

        We infer the mean wait time $\tau$ at each node from the mean shopping time of 5.95 minutes for small stores as reported in \cite{sorensen2017fundamental} and the mean shopping path length of 29.66 nodes based on the synthetic data described above.
        This yields a mean wait time of $\tau$ equals to $5.95 / 29.66 = \meanWaitTime$ minutes per node.
        Note that data on the mean shopping length and the arrival rate is from before the COVID-19 pandemic, so these values may be different during the pandemic due to a change in customer shopping behaviour.

        We use $p_I = 0.11\%$ as the proportion of infectious customers based on data reported from UK's Office for National Statistics (ONS) Infection Survey from September 2020 \cite{onsinfectionsurvey}.
        In this survey, a random sample of the population is tested for COVID-19 to estimate the overall proportion of people who have COVID-19 at that particular point in time.

        The transmission rate $\beta$ is much harder to estimate than the previous parameters because very little data exists.
        In \cite{tang2020}, the mean probability of transmission per contact is estimated to be $2.11 \times 10^{-8}$.
        However, they do not specify the duration of a contact.
        We use this parameter and assume that the mean contact duration is 15 minutes to obtain a rate of transmission to be $\beta = 2.11 \times 10^{-8} /15 = \betaVal$ per minute.
        (We assume that the probability of transmission is proportional to the contact duration.)

        We summarize the parameter values that we use in Table~\ref{tab:params}.

        \begin{table}[]
            \centering
            \caption{Parameter values that we use in our agent-based model}
            \label{tab:params}
            \begin{tabular*}{\textwidth}{| l @{\extracolsep{\fill}} |l |l | l|}
                \hline
                \textbf{Parameter} & \textbf{Default value} & \textbf{Reference/Assumption} \\ \hline
                Arrival rate ($\lambda$) & $2.55$ customer/min & \cite{Ying2019} \\ \hline
                Mean wait time at each node ($\tau$) & $\meanWaitTime$ min & inferred from \cite{sorensen2017fundamental} \\ \hline
                Percentage of infectious & $\pIVal$ & \cite{onsinfectionsurvey} \\
                 customers ($p_I$) & & \\ \hline
                Transmission rate ($\beta$) & $\betaVal$ per min & inferred from \cite{tang2020}   \\
                & & (assuming mean contact \\ 
                & & duration of 15 mins)\\ \hline
                Length of opening hours ($H$) & 14 hours & \cite{thesunsupermarketopeningtimes} \\ \hline
                Store layout & fixed & Layout of synthetic store\\\hline
            \end{tabular*}
        \end{table}

    \subsection*{Analysis environment}
        We implemented our agent-based model 3 in Python 3.6 using SimPy 4 \cite{simpy}.
        We ran our simulations on an Amazon Web Service cluster (ml.c5.4xlarge with 32 GB RAM and 16 vCPU) \cite{aws}, although all simulations can be performed on a standard desktop computer.

\section*{Results and discussion}
    \label{sec:results_and_discussion}

    \newcommand{\RRR}{\ensuremath{0.17}}
    \newcommand{\meanNumInfections}{\ensuremath{8.91\times 10^{-9}}}
    \newcommand{\meanNumInfectionsRRR}{\ensuremath{1.52\times 10^{-9}}}
    \newcommand{\chanceInfection}{\ensuremath{4.16\times 10^{-12}}}
    \newcommand{\meanShoppingTime}{\ensuremath{5.94}}
    \newcommand{\meanNumCust}{\ensuremath{14.90}}
    \newcommand{\totalExposureTime}{\ensuremath{6.32}}
    \newcommand{\totalExposureTimePerCust}{\ensuremath{0.0030}}
    \newcommand{\totalExposureTimePerCustSec}{\ensuremath{0.18}}

    We demonstrate in this section how to use our model, what metrics we can record in it, and what results one might obtain. 

    Using the default parameters listed in Table~\ref{tab:params}, we perform 1000 simulations, each simulating a day in our synthetic store.
    Customers stay on average $\meanShoppingTime$ minutes in the store, with on average $\meanNumCust$ customers present in the store at any given time.
    With $p_{I} = \pIVal$ infected customers, the total exposure time is on average $\totalExposureTime$ min per day.
    Multiplying this with $\beta = \betaVal$ infections per minute of exposure time, we estimate an average of $\meanNumInfections$ infections per day.
    Each susceptible customer is exposed $\totalExposureTimePerCust$ minutes (or about $\totalExposureTimePerCustSec$ seconds) on average per visit to an infected customer and the estimated a chance of infection is $\chanceInfection$.
    We list all metrics that we record in our simulations in Table~\ref{tab:metrics}.

    \begin{table}[!ht]
    \caption{Simulation results from 1000 simulations. 
    We show the mean and standard deviation of each metric across 1000 simulations.}
    \label{tab:metrics}
    \centering
    \begin{tabular}{|l |c |c |}
    \hline
    \textbf{Metric} & \textbf{Mean} & \textbf{Standard deviation} \\
    \hline    
     
    Number of customers & 2142 & 48.02  \\ \hline
    Number of susceptible customers & 2139 & 48.04 \\ \hline
    Number of infected customers & 2.56 & 1.67 \\ \hline
    Mean number of customers in store & $\meanNumCust$ & 0.44 \\ \hline
    Mean shopping time (in min) & $\meanShoppingTime$ & 0.06 \\ \hline
    Total exposure time (in min) & $\totalExposureTime$ & 4.62 \\ \hline
    Total exposure time per sus. customer (in min) & $\totalExposureTimePerCust$ & 0.0022 \\ \hline
    Number of infections & $\meanNumInfections$ & $6.52\times 10^{-9}$ \\ \hline
    Chance of infection per sus. customer & $\chanceInfection$ & $3.04\times 10^{-12}$ \\ \hline

    \end{tabular}
    \end{table}

\begin{figure}[htb]
        \centering
        \includegraphics[width=\textwidth,trim={5cm 1cm 9cm 0cm},clip]{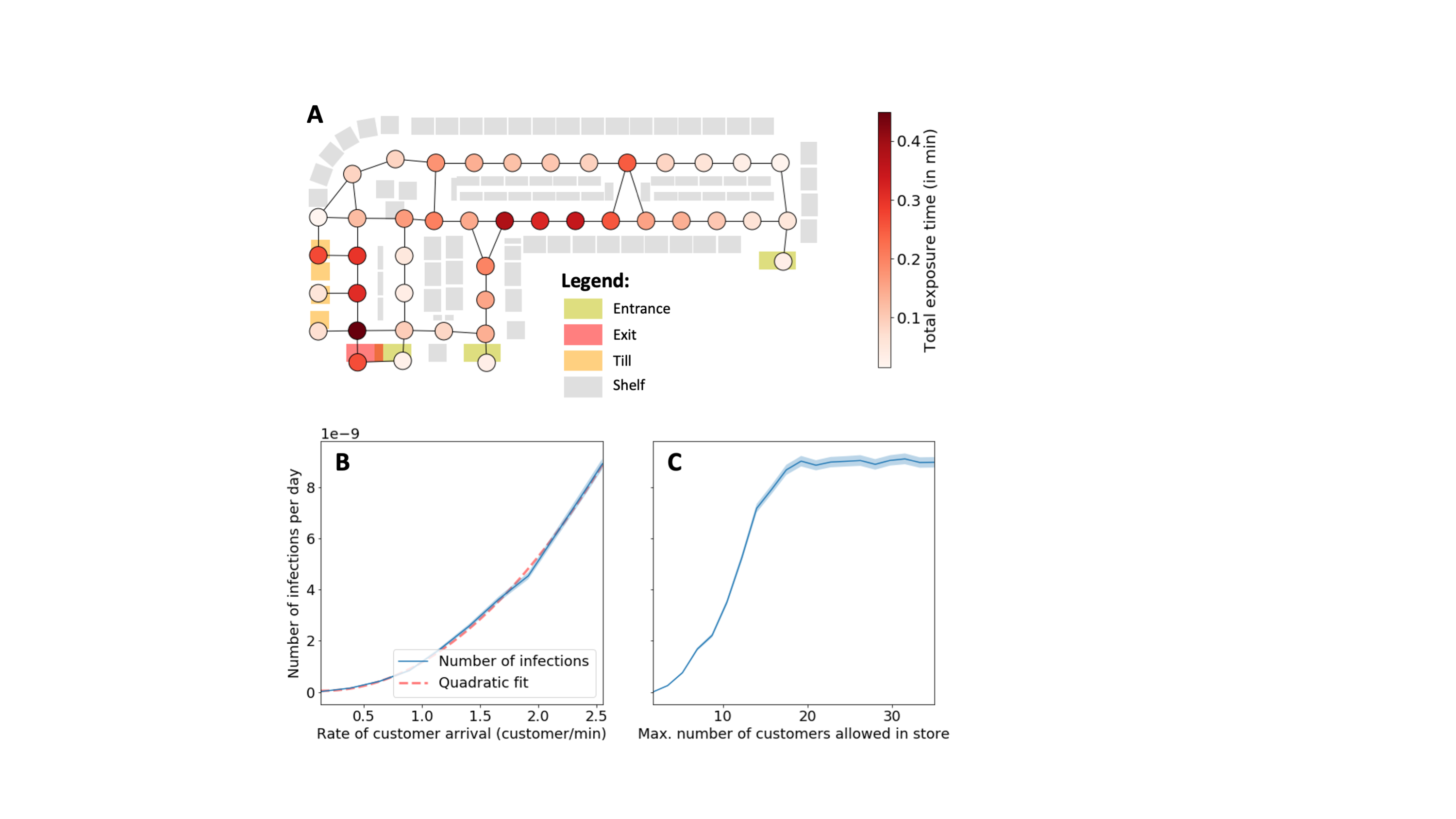}
        \caption{\textbf{(A)} Total exposure time per node. Nodes in the centre and near the tills of the store show significantly higher amount of exposure time than others.
        \textbf{(B + C)} The mean number of infections (with standard error) in a store across 1000 simulations as a function of the customer arrival time and maximum number $C_{\max}$ of customers (respectively). The mean number of infections scales quadratically with the arrival rate. In subfigure C, the mean number of infections plateaus, as the number of customers typically does not exceed 20 in our simulations.}
        \label{fig:initial_graphs}
    \end{figure}

    Our models also allow us to record the exposure time for each node. 
    We see in Fig~\ref{fig:initial_graphs}A that near the tills and centre of the store shows the highest amount of exposure time, thereby revealing mobility flow bottlenecks in the store that may be mitigated by changing to a different layout. 

    \subsection*{Varying customer arrival rate}

        Some stores restrict the rate at which customers enter the store. 
        We can incorporate this in our model by varying the arrival rate.
        We see that the number of infections show a quadratic dependence to the arrival rate (see Fig~\ref{fig:initial_graphs}B).

    \subsection*{Restricting maximum number of customers in store}
        Instead of directly reducing the rate of customer arrival, some stores opt to restrict the maximum number $C_{\max}$ of customers in a store.
        We can add this restriction to our model by simulating a queue outside of the store, where customers queue up if we have $C_{\max}$ or more customers in the store.
        Customers from the queue only enter the store when the number of customers in the store is below $C_{\max}$.
        In our model, the estimated number of infections also decreases significantly when decreasing the maximum number of customers in the store (see Fig~\ref{fig:initial_graphs}C).

    \subsection*{Face masks}

        Similar to \cite{li2020mask}, we model the implementation of a face mask policy via a reduction in the transmission rate.
        For example, \cite{chu2020physical} estimated the relative transmission risk reduction to be $RRR = \RRR$.
        We incorporate this by multiplying $\beta$ with $RRR$, reducing the number of infections by the same factor from $\meanNumInfections$ to $\meanNumInfectionsRRR$.

    \subsection*{One-way aisle layout}
\begin{figure}[!htb]
            \centering
            \includegraphics[width=\textwidth,trim={6cm 1cm 7cm 2cm},clip]{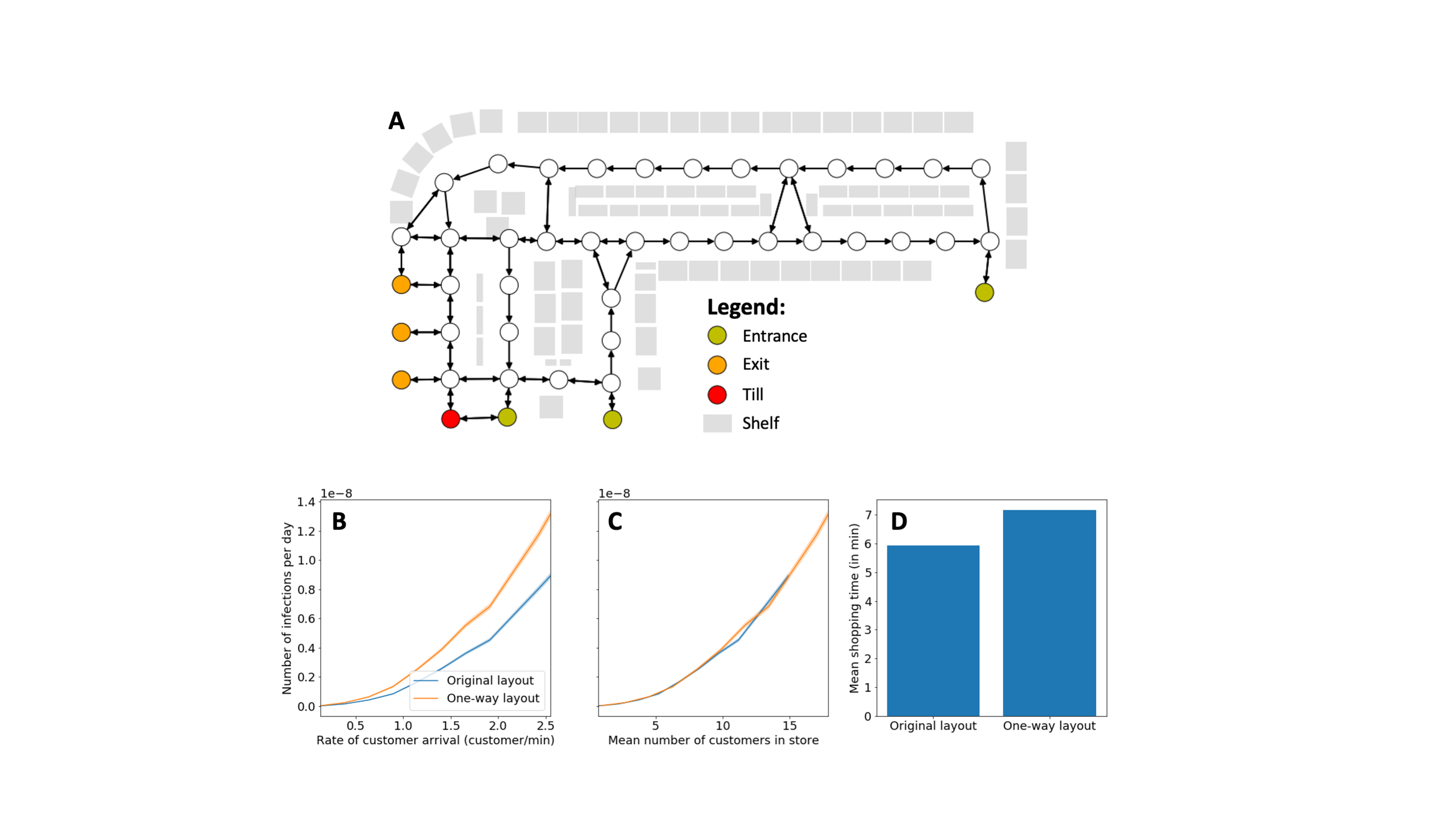}
            \caption{\textbf{(A)} Store layout with one-way aisles. 
            \textbf{(B + C)} Number of infections in a store as a function of the customer arrival time and mean number of customers (respectively).
            \textbf{(D)} Mean customer shopping time.} 
            \label{fig:oneway_graphs}
        \end{figure}

        A number of stores have implemented one-way systems to assist with social distancing and potentially redistributing the flow of customers.
        We can also assess this policy in our framework by changing the store graph to a directed graph, where some edges are uni-directional.
        For the synthetic store, we show an example one-way aisle layout in Fig~\ref{fig:oneway_graphs}A which we call the \emph{one-way store layout}.
        We need to change the shopping paths in our data, as they may violate the uni-directionality of the one-way store layout.
        For each path, we first consider again the node sequence $(v_1, \dots, v_k)$ from which the path was generated. (As a reminder, $v_1$ is an entrance node, $v_k$ is an exit node, and the intermediate nodes $v_2, \dots, v_{k-1}$ are locations where customers bought one or more items.)
        The corresponding path for the one-way store layout is then a shortest path that visits each of the nodes $v_1,\dots, v_{k-1}$ in sequence in the one-way store layout.

        According to our model, the one-way layout increases the number of infections (see Fig~\ref{fig:oneway_graphs}B).
        To explain this phenomenon, we can look at the mean shopping time between the two layouts.
        In the one-way store layout, the shopping path necessarily is at least as long as in the original store layout.
        Therefore, on average, the mean shopping time increases (see Fig~\ref{fig:oneway_graphs}D), so more customers are in the store at the same time and thus more infections occur.
        When we fix the mean number of customers in the store, the number of infections is largely the same (see Fig~\ref{fig:oneway_graphs}C).

    \subsection*{Limitations}
        Our model comes with a number of limitations.
        Firstly, a customer's path in our model does not depend on other customers, whereas we expect a customer's path to change depending on, for example, the crowdedness of other zones.
        Unlike in \cite{plata2020simulating}, our spatial topology is more coarsely grained, where customers are either in the same zone or not, and we therefore do not take the precise distance between customers into account.
        We also assume that the chance of infection is proportional to the exposure time, whereas in reality it may be non-linear (\eg, represented by a logistic function to model infectious dosage).
        In our simulations, we used a constant arrival rate and random shopping paths that do not change with time, while we expect time-varying arrival rates and shopping path distributions in reality.
        However, it is relatively straightforward to incorporate modifications to the transmission function or to the arrival rate into our agent-based model.
        Lastly, all results that we present in this article are from simulations on a synthetic data set and with a large uncertainty on the value of $\beta$. 
        Therefore the concrete results that we presented may have limited generalisation power. 
        The number of infections of store that we estimated should therefore not be taken at face value.
        Nonetheless, even without an accurate measure of $\beta$, we anticipate that the exposure time is a relevant metric to measure the relative risk of transmission.

\section*{Conclusion}

    We presented a model for modelling virus transmission (in particular, SARS-CoV-2 transmissions, which causes COVID-19, but it is more generally applicable) in supermarkets based on an agent-based model of customers traversing from zone to zone and being exposed to potential virus infection when in the same zone as an infected customer.
    We measured the risk of virus transmission by the total time that susceptible customers spent in the same zone as infected customers (and called this time the total exposure time).
    We demonstrated the capabilities of the model by applying it to synthetic data.
    We showed how one can use the model to identify hotspots and bottlenecks, estimate the reduction in virus transmissions in different scenarios such as restricting the maximum number of customers in a store or implementing a one-way aisle system.
    In particular, in our synthetic store, a directed store layout does not help in reducing the exposure time, as it increases the customer shopping time, so more customers are in the store at any given time.
    We invite retailers to use our models to identify bottlenecks that lead to crowded zones as well as to inform them on the best store policy.

\section*{Acknowledgments}

    This work was undertaken as a contribution to the Rapid Assistance in Modelling the Pandemic (RAMP) initiative, coordinated by the Royal Society.
    We thank David Romano-Critchley and Sumanas Sarma from Sainsbury's for providing the synthetic data.
    We also thank Mason A. Porter, Sam D. Howison, Mariano Beguerisse-D\'{i}az, John Fitzgerald, Mattie Landman, Mike Batty, and Philip Wilkinson for helpful discussions.

\nolinenumbers

%
%
%


\end{document}